\documentclass{ws-procs11x85}
\usepackage{ws-procs-thm}           
\usepackage{xcolor}

\usepackage{algorithm}
\usepackage{algpseudocode}
\usepackage{amsmath}
\usepackage{amsfonts}
\usepackage{booktabs}
\usepackage{graphicx}

\usepackage{color}

\DeclareMathOperator*{\argmin}{arg\,min}

\begin{document}

\title{Enhancing Model Interpretability and Accuracy for Disease Progression Prediction via Phenotype-Based Patient Similarity Learning}

\author{Yue Wang$^1$, Tong Wu$^{1,2}$, Yunlong Wang$^{1\dag}$, Gao Wang$^{3}$}

\address{
$^1$Advanced Analytics, IQIVA Inc., PA, USA\\
$^2$Biomedical Engineering, University of Minnesota, MN, USA\\
$^3$Department of Human Genetics, University of Chicago, IL, USA\\
$^\dag$E-mail: Yunlong.Wang@iqvia.com}

\begin{abstract}
Models have been proposed to extract temporal patterns from longitudinal electronic health records (EHR) for clinical predictive models. 
However, the common relations among patients (e.g., receiving the same medical treatments) were rarely considered.
In this paper, we propose to learn patient similarity features as phenotypes from the aggregated patient-medical service matrix using non-negative matrix factorization.
On real-world medical claim data, we show that the learned phenotypes are coherent within each group, and also explanatory and indicative of targeted diseases.
We conducted experiments to predict the diagnoses for Chronic Lymphocytic Leukemia (CLL) patients. 
Results show that the phenotype-based similarity features can improve prediction over multiple baselines, including logistic regression, random forest, convolutional neural network, and more.
\end{abstract}


\copyrightinfo{\copyright\ 2019 The Authors. Open Access chapter published by World Scientific Publishing Company and distributed under the terms of the Creative Commons Attribution Non-Commercial (CC BY-NC) 4.0 License.}

\section{Introduction}

Recent years have witnessed an explosion in the growth of electronic health records (EHR), which has motivated the use of machine learning, especially deep learning methods in disease prediction \cite{choi2016using,yu2019rare,lipton2015learning,choi2016retain} and disease progression modeling \cite{baytas2017patient,che2017rnn}.
To improve the prediction performance of these tasks, as pointed out by previous studies ~\cite{baytas2017patient,che2017rnn}, many diseases demonstrate heterogeneous progression patterns. 
A preferred strategy  to resolve this heterogeneity is to segment out more homogeneous
patient sub-populations based on patient similarity.



There have been a few attempts in recent years that exploit patient similarity to improve disease progression modeling.
For example, Che \emph{et al.} \cite{che2017rnn} and Baytas \emph{et al.} \cite{baytas2017patient} proposed RNN based models to compute patient similarity based on their longitudinal medical records for improved personalized disease prediction. Fan \emph{et al.} leveraged the relational structure between patients and doctors by modeling it as a bipartite graph, and extracted similarity features of patients from the bipartite graph using the method graph Laplacian to improve the accuracy of predicting first-line treatment initiation \cite{zhang2019predicting}.

Despite the initial success, these models often solely rely on patient records, while ignoring the structural information among patient population encoded in the EHR systems. e.g., similar patients often share a similar set of medical services. We postulate that these structural information bears a direct and strong indication of patients' disease progression.
However, there is a caveat that concerns of directly including these information into similarity measurements. For example, routine services (e.g., measuring blood counts) can constitute a significant portion of the occurrences of all medical services in patient journeys, and would ``overshadow'' other infrequent and more complicated services due to their extremely imbalanced occurrences in patient journeys.

We have made the following contributions to solve the aforementioned challenges.

\begin{itemize}
\item We extract phenotypes that encapsulate related medical services as a clinically meaningful set from patient journeys.
Each phenotype refers to a collection of clinical traits and measurements that describe fundamental attributes of a patient \cite{frey2014ehr}.
As a result, patient similarity can be measured on the level of phenotypes instead of individual services, thereby preventing complicated medical services from being ``overshadowed'' by the large amount of routine services.
\item Phenotypic structure of medical services can also enhance the interpretability of computational models that leverage such information to better portrait the clinical relevance of patients.
In addition, patient phenotyping allows straightforward inference of vital information of patient profiles even in the case of missing data, which is pervasive in clinical settings \cite{beaulieu2018characterizing}.
For example, we can infer phenotypes of patients with missing data by projecting their medical records onto the space defined by the learned phenotypic feature vectors.
\item We propose a Wide \& Deep neural network model that can effectively combine latent knowledge learned from both patients' longitudinal medical records and structural information (e.g., phenotype-based similarity features) to make more accurate predictions.
The model structure is inspired by the Wide \& Deep framework \cite{cheng2016wide} proposed originally for mobile app recommendations.
\end{itemize}

We demonstrate the effectiveness of the phenotype-based similarity features by predicting the diagnoses of chronic lymphocytic leukemia (CLL) patients.
Experimental results show that leveraging the combination of advanced machine learning algorithms and patients' medical footprints can allow us to exploit explainable and indicative phenotypes of CLL, which can provide better accuracy in diagnosis prediction than using conventional EHR data alone without the phenotypes on a variety choice of prediction models, including the Wide \& Deep model.
Furthermore, a more in-depth understanding of such disease phenotypes on the population level can aid doctors to detect easily-missed-diagnosis diseases in an early stage and reduce potential medical burden. 





\section{Methods}


\subsection{Problem Definition}\label{aba:sec1}

We let $\Omega$ denote the clinical medical records of all studied $P$ patients.
The medical record of a patient $p$ is $\Omega^{(p)}=\{\mathbf{S}^{(p)}, \mathcal{I}^{(p)}\}$, where $\mathbf{S}^{(p)}=[\mathbf{s}_1^{(p)},\mathbf{s}_2^{(p)},\dots,\mathbf{s}_T^{(p)}]$ is the sequence of medical services the patient has received and $\mathcal{I}^{(p)}$ represents the patient's demographics and other non-sequential features.
Each service $\mathbf{s}_t^{(p)}$ indicates one unique type of diagnoses, procedures, or prescriptions.

We denote as $T$ (in the unit of week) the length of patient journeys that we use in this paper.
We bin the counts of every unique service a patient has received in each week, and eventually obtain a $\mathbf{X}^{(p)} \in \mathbb{R}^{|\mathcal{S}| \times T}$ matrix for each patient, where $|\mathcal{S}|$ denotes the total number of unique services in the cohort, and $\mathbf{X}^{(p)}[i,t]$ denotes the counts of service $i$ the patient $p$ has received in week $t$.


To allow for phenotyping of medical services, we sum the matrix $\mathbf{X}^{(p)}$ along the temporal axis and thus ``fold'' it into a vector $\mathbf{x}_p \in \mathbb{R}^{|\mathcal{S}|}$, which summarizes the counts of all unique services that patient $p$ has received.
Next, we stack all $\mathbf{x}_p$ into a new matrix $\mathbf{Y} \in \mathbb{R}^{P \times |\mathcal{S}|}$, and perform non-negative matrix factorization (NMF) on $\mathbf{Y}$ to obtain a low-rank approximation, thereby unveiling a phenotypic structure of all services as well as the contribution of each phenotype to the patient journeys for each patient.
The extracted features from NMF will be included into $\Omega^{(p)}$ for model learning and prediction.

The goal of prediction is to learn $f:\Omega^{(p)} \mapsto y^{(p)}$, where $y^{(p)}$ is the label indicating if the patient is diagnosed with CLL.
The model outputs $Pr(y^{(p)}=1|\Omega^{(p)}) = f(\Omega^{(p)})$ that evaluates the likelihood of patient $p$ being diagnosed with CLL.

\subsection{Non-negative Matrix Factorization}

Given a non-negative matrix $\mathbf{Y} \in \mathbb{R}^{P \times |\mathcal{S}|}$, where $\mathbb{R}$ denotes the set of all real numbers,
we are interested in finding non-negative matrix factors $\mathbf{W} \in \mathbb{R}^{P \times r}$ and $\mathbf{H} \in \mathbb{R}^{r \times |\mathcal{S}|}$ with $r \ll |\mathcal{S}|$ such that
\begin{equation}
    \mathbf{Y} \approx \mathbf{W} \cdot \mathbf{H},
\end{equation}
where $\mathbf{W}$ and $\mathbf{H}$ can be obtained by solving the following optimization problem:
\begin{equation}
    \centering
    \begin{aligned}
        \mathbf{W}^{\star},\; & \mathbf{H}^{\star} = \argmin_{\mathbf{W, H}} ||\mathbf{Y}-\mathbf{WH}||_2^2, \\
        &\text{subject to} \; \mathbf{W,H} \geq 0.
    \end{aligned}    
\end{equation}
We can use the \emph{multiplicative update rules} \cite{lee2001algorithms} to calculate the values of $\mathbf{W}$ and $\mathbf{H}$ as:
\begin{align}
    \mathbf{H}_{i,j}^{n+1} &\gets \mathbf{H}_{i,j}^n\frac{((\mathbf{W}^n)^T\mathbf{Y})_{i,j}}{((\mathbf{W}^n)^T\mathbf{W}^n\mathbf{H}^n)_{i,j}} \\
    \mathbf{W}_{i,j}^{n+1} &\gets \mathbf{W}_{i,j}^n\frac{(\mathbf{Y}(\mathbf{H}^{n+1})^T)_{i,j}}{(\mathbf{W}^n\mathbf{H}^{n+1}(\mathbf{H}^{n+1})^T)_{i,j}},
\end{align}
where $n$ denotes the index of the iteration.

After NMF, the columns of $\mathbf{H}$ correspond to all the medical services appeared in the patient journeys.
In each row of $\mathbf{H}$, the majority of the nonzero terms constitute a clinically meaningful set of medical services, or phenotypes, such as oncological conditions or neurological system disorders.
In other words, $\mathbf{H}$ defines a set of axes over the entire set of medical services, where each medical service resides in one of the axes and has minor influence on the rest.
Meanwhile, the rows of $\mathbf{W}$ correspond to all the patients, and the $r$ elements in one row specify the assignments of the current patient to each of the $r$ medical phenotypes.
Therefore, we could characterize the disease phenotype of a patient by examining the corresponding row in the $\mathbf{W}$ matrix, which can be used to enhance the model interpretability.

In the training data set, we have both CLL diagnosed patients as positive and undiagnosed patients as negative, for which the $y^{(p)}=1$ or $0$ respectively. The cohort definition with specific clinical rules will be given in section \ref{results:sec1}. 
To obtain phenotypes of medical services with better differentiability and relevance associated with CLL, we run NMF only on $\mathbf{Y}_{pos}$, i.e., the patient-service matrix derived from the patients who are diagnosed with CLL (the positive cohort), and obtain their phenotype vectors $\mathbf{W}_{pos}$; for the patients in the negative cohort, we project their patient-service matrix $\mathbf{Y}_{neg}$ onto the phenotype axes $\mathbf{H}_{pos}$ obtained from $\mathbf{Y}_{pos}$ via non-negative least squares. Specifically, for each row vector $\mathbf{w}_i \in \mathbb{R}^r$ of $\mathbf{W}_{neg}$, we got the least square estimate by solving the following optimization problem:
\begin{equation}
    \centering
    \begin{aligned}
    \argmin_{\mathbf{w}_i} ||\mathbf{Y}_{neg}^T - \mathbf{H}_{pos}^T\mathbf{w}_i^T ||_2^2 \\
    \text{subjective to} \: \mathbf{w}_i \geq 0,
    \end{aligned}
\end{equation}
after which we concatenate all $\mathbf{w}_i$ into $\mathbf{W}_{neg}$.

\subsection{Wide and Deep}

In the original Wide \& Deep framework\cite{cheng2016wide}, the wide component \emph{memorizes} frequent co-occurrence of input features, and the deep component \emph{generalizes} or \emph{explores} new feature combinations that rarely occurred by first projecting input features into low-dimensional embeddings and exploring their interactions via deep neural networks.
Wide \& Deep provides a scalable and effective way of balancing memorization and generalization of input features and their interactions, and makes better recommendations.

In this work, we make a key modification to the deep component by replacing it with an LSTM to account for the sequential information embedded in the patients' longitudinal medical records.
We also deploy a pre-trained embedding layer at the input of the LSTM to project high-dimensional and sparse medical services into low-dimensional dense vectors, which facilitates efficient exploration of the complex dependencies and nonlinear dynamics of patients' medical services.
The embeddings of medical services were trained using Word2Vec \cite{mikolov2013distributed} by maximizing the likelihood of each service seeing its neighbors in the patient journeys.

\vspace{5pt}
\noindent
\textbf{The wide component.} 
The inputs to the wide component include patients' demographics and disease phenotypes extracted using NMF from the patient-service matrix $\mathbf{Y}$.
We encode the patients' demographics as categorical variables, e.g., $\mathbf{f}_1$, $\mathbf{f}_2$, ..., $\mathbf{f}_N$, where each $\mathbf{f}_i \in \{0,1\}^{|\mathbf{f}_i|}$ with only one 1 out of $|\mathbf{f}_i|$ bits that denotes a distinct value of feature $\mathbf{f}_i$.
The phenotypic feature for patient $p$ is a vector $\mathbf{w}_p \in \mathbb{R}^{r}$.
The value of $r$ is set as 4 empirically.

The phenotypic features and the one-hot encoded demographics are then incorporated into the overall prediction through three consecutive dense layers along with features extracted by the deep component, which will be introduced next.

\vspace{5pt}
\noindent
\textbf{The deep component.}
In the original Wide \& Deep model for recommendation \cite{cheng2016wide}, the deep component is implemented as several stacked fully-connected layers.
For the purpose of predicting diagnosis, a sequential model that can account for the temporal progression of time-stamped medical records, captures more representative information.

In this work, we adopt LSTM as the primary structure in the deep component to model the sequential information embedded in the patient journeys.
The inputs to the deep component consist of sequences of medical services.
Before entering the LSTM, all medical services are converted into $d$-dimensional dense vectors through an embedding layer.
Representing medical services as low-dimensional embedding vectors instead of one-hot codes can better preserve the structural relations of medical services, thereby improving model performance that take embedding vectors as inputs.
For service $i$, its embedding vector is:
\begin{equation}
    \mathbf{s}_i=\text{Embedding}(s_i) \quad \text{for} \; i=1 \cdots |\mathcal{S}|,
\end{equation}
where $\mathbf{s}_i \in \mathbb{R}^d$.

At each time step of the LSTM, we apply the \texttt{GlobalMaxPooling} operation to extract the most significant hidden features:
\begin{align}
    \mathbf{h}_1, \mathbf{h}_2, \dots, \mathbf{h}_T &= \text{LSTM}(\mathbf{s}_1, \mathbf{s}_2, \dots, \mathbf{s}_T), \\
    \mathbf{h}_{max} &= \text{GlobalMaxPooling}([\mathbf{h}_1, \mathbf{h}_2, \dots, \mathbf{h}_T])
\end{align}
where $\mathbf{h}_t \in \mathbf{R}^{d'}$ ($d'$ is the dimension of the hidden variable of the LSTM), $\mathbf{h}_{max} \in \mathbb{R}^T$;
the subscript of service vector $s$ is the index of time, not the service type.
Next, we concatenate $\mathbf{h}_{max}$ with the features from the wide component, and feed them through three dense layers which fully explore the interaction between the wide and deep features:
\begin{equation}
    \text{P}(y^{(p)}=1) = \sigma(\mathbf{W}_{3}^T \cdot \text{ReLU}(\mathbf{W}_{2}^T \cdot \text{ReLU}(\mathbf{W}_{1}^T\cdot[\mathbf{h}_{max};\mathbf{w}_p;\mathbf{f}_{1:N}^{(p)}]+\mathbf{b}_{1})+\mathbf{b}_{2}) + \mathbf{b}_{3}).
\end{equation}

\section{Results}

\subsection{Cohort Selection and Data Extraction}
\label{results:sec1}

CLL is the most common type of leukemia in adults.
However, many CLL patients do not have obvious symptoms before CLL is diagnosed.
The leukemia is often found when their doctor orders blood tests for some unrelated health problem or during a routine check-up and they are found to have a high number of lymphocytes.
Common signs involving fatigue, weight loss, and fever are not reliable indicators to raise a doctor's awareness of severe medical conditions. 
Only until the blood test performed, CLL is hard to detect and diagnose. 

We extract data from IQVIA longitudinal prescription (Rx) and medical claims (Dx) database, including hundreds of millions of patient clinical records.
We select patients diagnosed with CLL from January 2012 to December 2018 from the IQVIA database and keep only the patients with complete Rx/Dx information as the positive cohort.
For each patient in the positive cohort, we pull one-year clinical records six months before the date of the first CLL diagnosis to avoid any potential information leakage.
To generate the negative cohort, we extract all the patients from January 2018 to December 2018 who have shown CLL risk factors and related symptoms but never been diagnosed. 
The risk factors are empirically selected from medical literature and will be specifically defined later. 
Patients are greater than 18 years old in both the positive and negative cohort.

The including criterion for generating the negative cohort contains four parts: 
CLL risk factor, CLL related diagnosis, CLL related procedure, and CLL related prescription. 
A negative patient in our study must meet at least three out of the four criteria. 
The risk factors include anemia, chills, fatigue, fever, night sweats and pain, Sj$\ddot{o}$gren syndrome, weakness, and weight loss. 
The related diagnoses include Epstein-Barr virus infection, recurrent infection, Helicobacter pylori infection, HIV/AIDS, Human T-lymphotrophic virus type I, hypogammaglobulinemia, psoriasis, rheumatoid arthritis, Wiskott-Aldrich syndrome, and pneumonia. 
The related procedures include increased frequency of CBC/ or blood test, tissue culture and chromosome analysis, and flow cytometry. 
The related prescriptions include Dexamethasone, Neupogen, and Prednisolone. 
By applying the inclusion and exclusion criteria, we identify 19,425 positive and 177,316 negative patients. 

\subsection{Features Extraction}

\noindent
\textbf{Demographics.}$\:$
Demographical features include age and gender, both of which are categorical variables. 
Gender has three categories that are male, female, and unknown. 
Age is categorized as equal to or greater than sixty, smaller than sixty, and unknown. 
The age and gender information is based on the index date of each patient that is the first diagnosis date for the positive cohort and December 31, 2018 for the negative cohort. 

\vspace{5pt}
\noindent
\textbf{Clinical service features.}$\:$
The prefix at the beginning of a service indicates the type of the service:
`dx-' stands for a diagnosis, `rx-' stands for a prescription, and `px-' stands for a procedure.
To reduce the number of unique services and improve the interpretability, we aggregate the clinical codes that essentially refer to the same service subtype into a clinical category.
For instance, we use the clinical category `dx-shigellosis' to represent a group of fourteen diagnosis codes. 
Furthermore, we ignore the clinical categories that have less than 5\% occurrences for all patients. 
The final model inputs include 394 unique clinical features. 

The patient journey is ordered by service date.
If a patient has multiple types of service on the same day, we order these types by procedure first, diagnosis second, and prescription last.
Services of the same type that happen on the same day will be ranked alphabetically.

\vspace{5pt}
\noindent
\textbf{Phenotyping medical services with NMF.}$\:$
We utilized the Bag-of-Word approach\cite{mikolov2013efficient} to tokenize each patient's service journey and transformed the clinical features into a sparse patient-service matrix (discussed in Section 2.2). 
To decide the number of phenotypes, we run multiple times of NMF with different numbers of phenotypes (i.e., the value of $r$ was tested from two to eight), and asked a medical expert to evaluate the quality of the phenotypes.
For the $i$-th phenotype ($1 \leq i \leq r$), it includes all the medical services of which the $i$-th coefficients are the maximum in their corresponding $r$-dimensional column vectors in $\mathbf{H}$.
If all the coefficients of a medical service after NMF are too small (less than 0.001 in our setting), this service will be assigned into a miscellaneous group.

Eventually, we found that the choice of three phenotypes when performing NMF can give the best differentiation between medical services.
We asked a medical expert to label each phenotype (excluding the miscellaneous group) with a general description according to the clinical meanings of its members.
The obtained three phenotypes are:
\begin{itemize}
\item{\textbf{Respiratory disease and pain (\#1):}} Disorders involving the respiratory system and general pain, including pneumonia, obstructive chronic bronchitis, disease of pleura, and chronic pain, etc. Respiratory tract illness is common in CLL patients, and the most common complication is pneumonia\cite{ahmed2003pulmonary}. Pulmonary involvement in patients with CLL may result from infections and pathological infiltration\cite{hill2012pulmonary}. 
General pain is also common in CLL patients, which depends on the location of the mass of abnormal cells. 
Chest pain and bone pain are common as well\cite{painfacts2012}.

\item{\textbf{Blood-related/anemia (\# 2):}} Clinical services related to blood, including coagulopathy test, chemotherapy, and diagnoses of anemia such as unspecified anemia and acquired hemolytic anemia. CLL is a type of blood cancer that impacts normal blood cells and bone marrow. The increased blood-related medical services or diagnosis implies the potential progression of CLL.

\item{\textbf{Infections (\#3):}} Diagnoses and procedures resulting from or related to infections that can cause a range of symptoms. Top-ranked clinical features include acute upper respiratory infections, intestine infections, and gastritis, etc. 
As a common complication of CLL, infection pertains to the alteration of immune system. 
Even in the early stages, nearly all CLL patients have decreased immunoglobulin levels that lead to infection\cite{dearden2008disease}.
\end{itemize}

\begin{figure}[t]
\begin{center}
\centerline{\includegraphics[width=0.9\columnwidth]{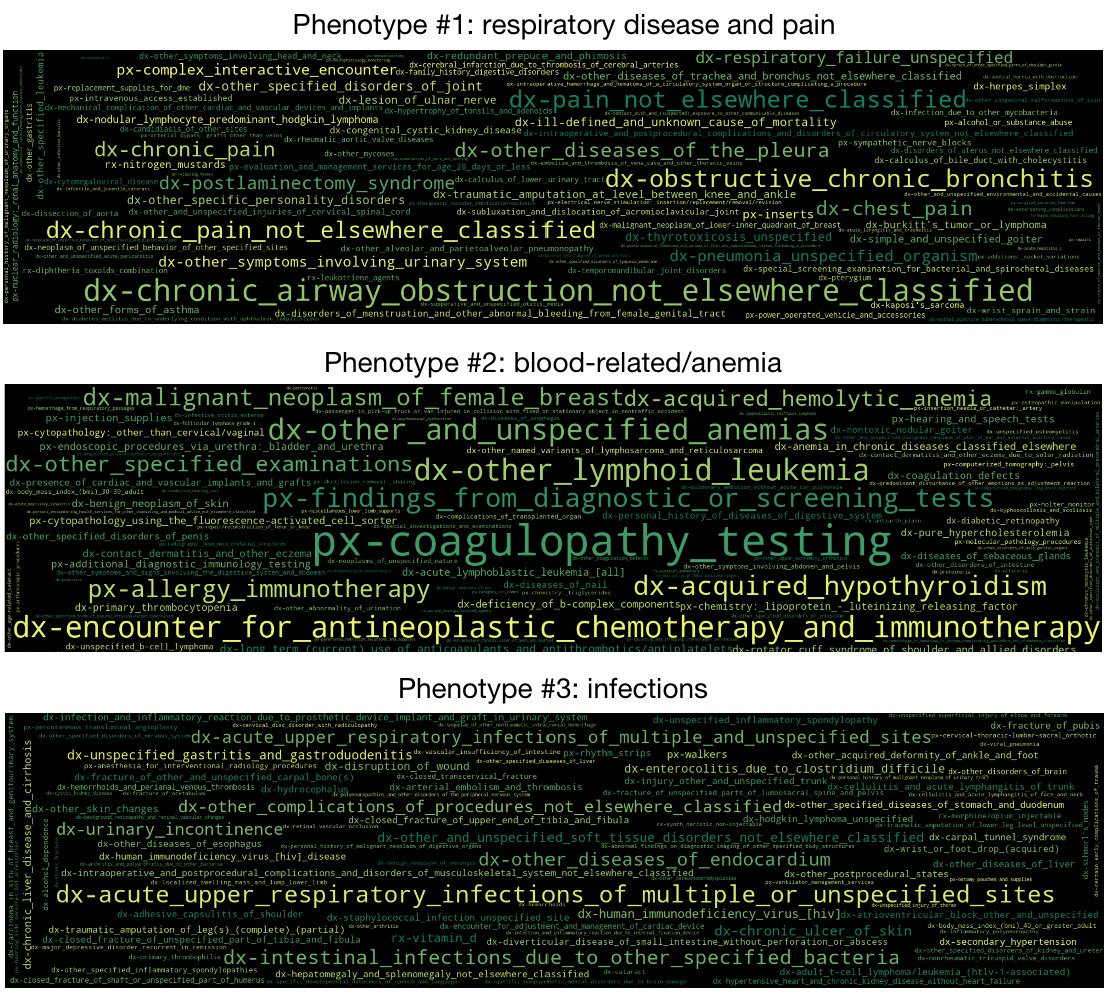}}
\caption{Word cloud of the three phenotype. In each subplot, the font size of a medical service is proportional to its coefficient that indicates its relevance to the corresponding phenotype.}
\label{wordc}
\end{center}
\vspace{-20pt}
\end{figure}

\begin{figure}[t]
\begin{center}
\centerline{\includegraphics[width=0.9\columnwidth]{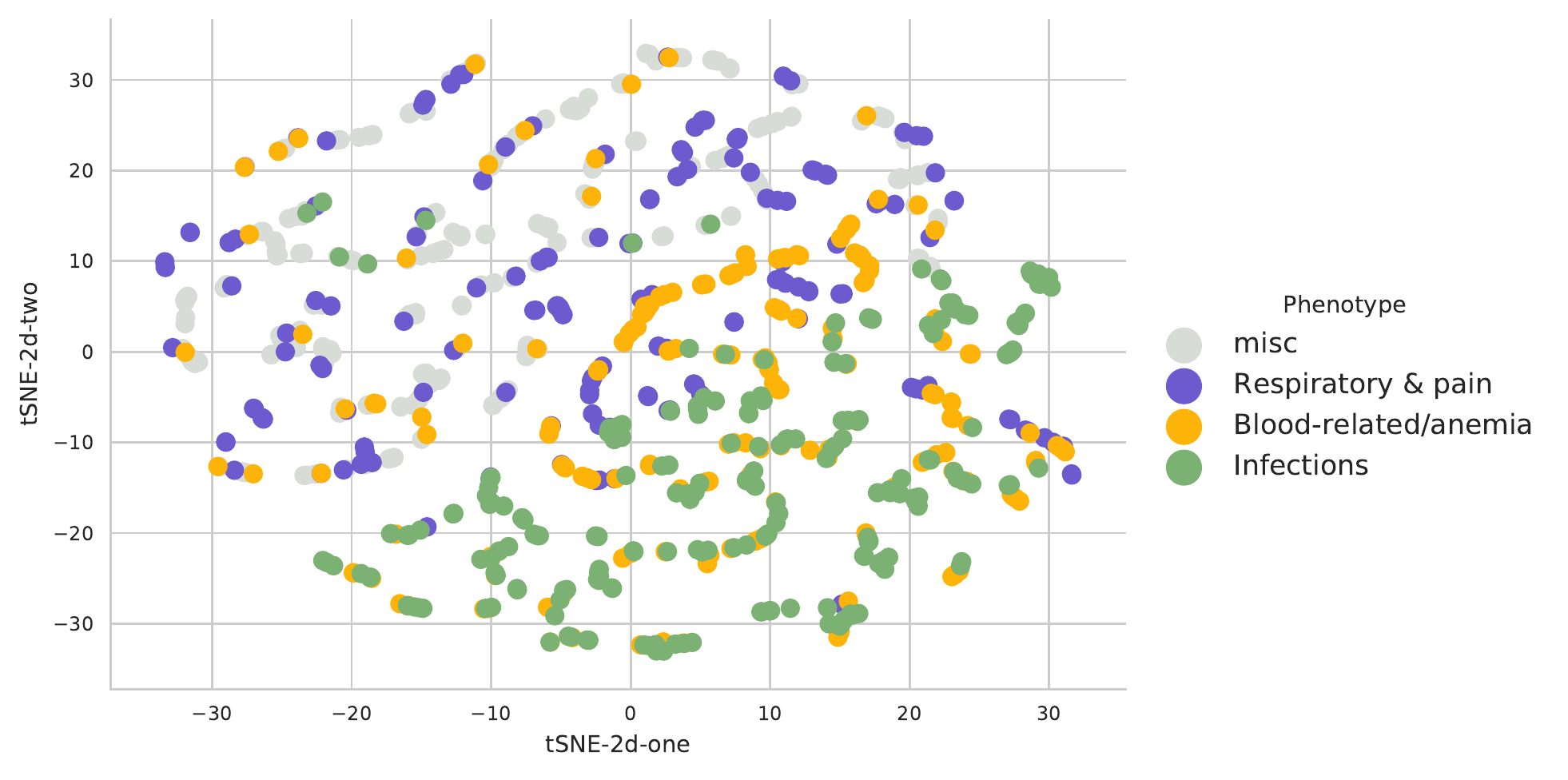}}
\caption{t-SNE visualization of medical services. The color of a medical service indicates its phenotype.}
\label{tsne}
\end{center}
\vspace{-20pt}
\end{figure}

Figure \ref{wordc} shows the top-ranked (the first 100) medical services from each phenotype in the form of word cloud, where the font size of a medical service is proportional to its NMF coefficient that denotes its belonging to a phenotype.
We further visualize the distribution of the medical services annotated with their phenotyping properties in the latent space defined by the NMF feature vectors.
To do that, we performed t-SNE \cite{maaten2008visualizing} on the $\mathbf{H}$ matrix that describes the phenotyping feature vectors of the medical services.
The 2-D scatter plot of the medical services after t-SNE is shown in Figure \ref{tsne}, where three groups of medical services encoded with different colors can be clearly identified.
The phenotype 2 (blood-related/anemia) partially overlaps with both the phenotype 1 (respiratory disease and pain) and the phenotype 3 (infections), respectively, whereas the phenotype 1 and 3 are clearly separated with minor overlap. 
The distributions of medical services among three groups align with our expectations. 
Essentially, the root cause of phenotype 2 is the disease progression of CLL. 
As elucidated in the phenotype description above, the abnormality in blood cells and bone marrow would lead to infections that are associated with respiratory tract inflammation.

\begin{figure}[t]
\begin{center}
\centerline{\includegraphics[width=0.6\columnwidth]{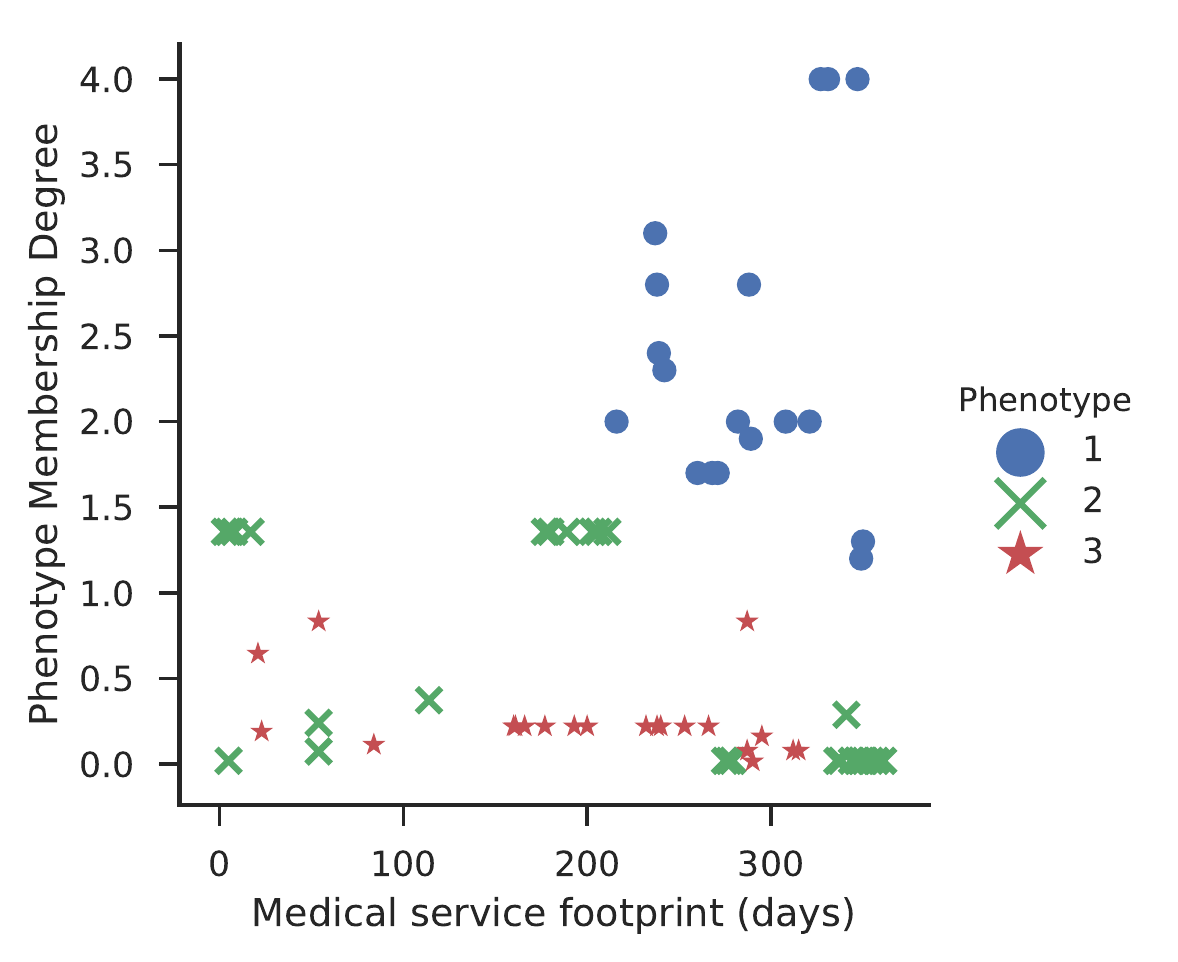}}
\caption{A patient showcase of fitted NMF phenotype. The X axis is the number of service days from the first day of the patient's look back period. The Y axis (Phenotype membership degree) shows the degree value of each service to the corresponding phenotype}
\label{pone}
\end{center}
\vspace{-15pt}
\end{figure}


\vspace{5pt}
\noindent
\textbf{Case study: a patient journey as phenotypes.}$\:$
In Figure \ref{pone}, we plot the temporal progression of a patient journey not in specific medical services, but phenotypes.
It shows clearly the disease progression wherein the patient has significantly increased services related to respiratory system and pain especially in the last one third of the look back period.
This illustrates how we can gain understandable and informative knowledge from patient-level data by leveraging the patient's phenotypic tendency.
Furthermore, it shows the NMF-based phenotype has the capability to capture and distinguish meaningful clinical services that suggest patients' disease progression.

\subsection{Model and Performance}
Because of the imbalanced positive and negative patients, we use precision-recall area-under-curve(PR-AUC) and F1 scores at various sensitivity values to evaluate model performances. 

To illustrate the effectiveness of NMF phenotype features, we test four different models:
logistic regression (LR), random forest (RF), convolutional neural network (CNN), and the modified Wide \& Deep.
The inputs to the tested models include the demographic and clinical service features, and with or without the phenotype features.

The train/test split ratio is 0.9:0.1. Within the training cohort, the train/validate split ration is 0.9:0.1.
The maximum training iteration of LR is 5,000.
Both CNN and Wide \& Deep are updated with the Adam optimizer\cite{kingma2014adam} with a default learning rate of 1e-4 and a batch size of 256.
The number of training epochs is 50;
we early stop the training if no improvements in five consecutive epochs.
All experiments were run in Keras with Tensorflow backend on a system with 128GB RAM, 16 Interl(R) Core Xeon(R) E5-2683 v4 2.10GHz CPUs, and Nvidia Tesla P100-PCIE-16GB.

\begin{figure}[t]
\begin{center}
\centerline{\includegraphics[width=0.95\columnwidth]{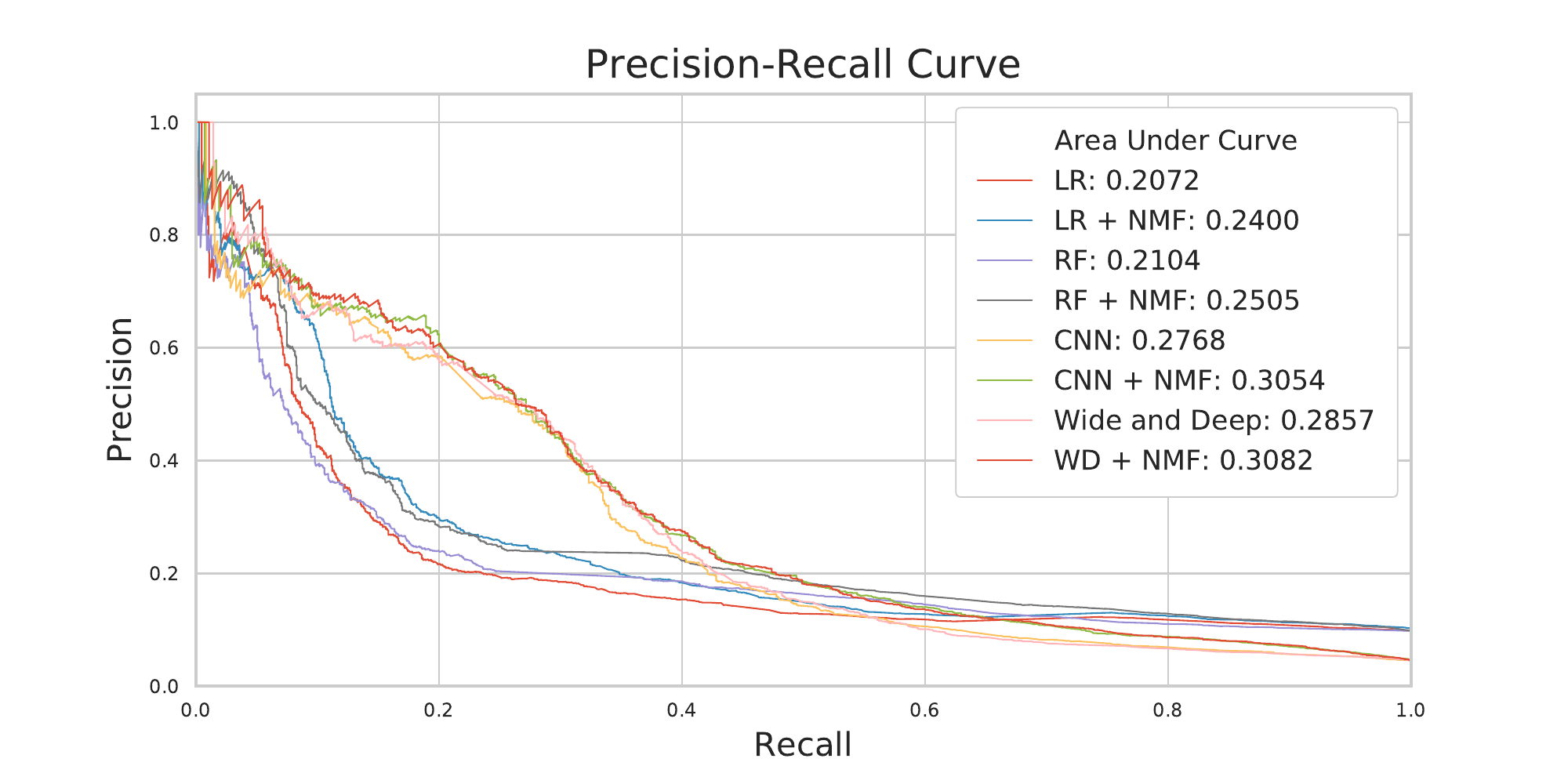}}
\caption{Comparison of baseline models with and without NMF phenotype feature measured in PR-AUC.}
\label{pr}
\end{center}
\vspace{-20pt}
\end{figure}

\vspace{5pt}
\noindent
\textbf{Model description.}$\:$
The RF model has 100 trees.
The CNN model employs a 4-layer structure $2\times2(128)-3\times3(128)-4\times4(128)-5\times5(128)$ (kernel size/number of kernels), followed by two dense layers. 
The flattened CNN features are concatenated with demographic and NMF features at the first dense layer. 
The Wide \& Deep model deploys an LSTM with 512 hidden dimension handling sequential medical features as the deep part, and a top-level two-layer feed-forward network to concatenate demographics and NMF features as the wide part. 


\vspace{5pt}
\noindent
\textbf{Performance.}$\:$
Overall, the PR-AUC and F1 score at various sensitivity levels provided in Figure \ref{pr} and Table \ref{pkkk} demonstrate that the models with NMF phenotype features outperform those without NMF phenotype features.
The margin on PR-AUC is more than 2\% on average across all model structures. 
The F1 score comparison reveals a similar pattern that the models with phenotype features in general have a higher F1 score at every sensitivity level. 
The results suggest the universal utility of phenotype features extracted with NMF from patients medical records in enhancing the performance of disease progression prediction.

\begin{table}[t]
\footnotesize
\centering
\caption{Comparison of F1 scores between proposed and benchmark methods}
\label{pkkk}
\begin{tabular}{l*{9}{c}r}
\hline
\hline
Sensitivity & $\textrm{LR}$ & $\textrm{LR+NMF}$ & $\textrm{RF}$ & $\textrm{RF+NMF}$ & $\textrm{CNN}$  & $\textrm{CNN+NMF}$ & $\textrm{WD}$ & $\textrm{WD+NMF}$\\
\hline
0.15            & 0.1974  & 0.2168 &  0.1997  & 0.2161  & 0.2433  & 0.2425 & 0.2392 & \bf 0.2483 \\
0.2           & 0.2079  & 0.2390 &  0.2085  & 0.2243  & 0.2973  & 0.3020 & 0.2955 & \bf 0.2976 \\
0.25            & 0.2174  & 0.2531 &  0.2156  & 0.2342  & 0.3365  & 0.3385 & 0.3370 & \bf 0.3381 \\
0.3           & 0.2289  & 0.2613 &  0.2234  & 0.2478  & 0.3551  & 0.3563 & 0.3568 & \bf 0.3579 \\
0.35            & 0.2236  & 0.2529 &  0.2534  & 0.2887  & 0.3148  & 0.3437 & 0.3441 & \bf 0.3433 \\
0.4           & 0.2224  & 0.2512 &  0.2513  & 0.2811  & 0.2891  & 0.3195 & 0.3009 & \bf 0.3237 \\
\hline
\end{tabular}

\label{pat_para_res}

\end{table}


\vspace{5pt}
\noindent
\textbf{Limitations.}$\:$
One limitation of the current work is that in the deep component we did not take into consideration the duration of time intervals between every two consecutive services.
Incorporating this time-span feature has a number of advantages:
First, predicting future time-spans can be included as an auxiliary task that acts as a regularization term in the loss function, which can potentially enhance the model's prediction accuracy \cite{choi2016doctor}.
Second, predicting time-spans naturally coincides with survival analysis, which offers the possibility of conducting survival analysis with deep neural networks \cite{ranganath2016deep}.

\vspace{5pt}
\noindent
\textbf{Future Works.}$\:$
First, we will extend the matrix factorization technique employed in this work to tensor decomposition by further taking the temporal dimension into consideration when building the aggregated patient-service high-dimensional tensor.
This will potentially offer more sufficient granularity and patient-specific characterizations of disease progression.
Second, to address the aforementioned limitation, we will investigate how to expand the current deep learning model structure to conduct survival analysis by, for example, explicitly modeling the time-span features that is currently missing from our implementation.

\section{Conclusion}
In this paper, we proposed to extract patient similarity features as phenotypes of medical services using non-negative matrix factorization.
We showed that the phenotype features learned automatically from patients medical records are highly explanatory and indicative of the targeted disease in the cohort.
We also proposed a modified version of the Wide \& Deep model that can effectively combine sequential modeling of patients' longitudinal clinical records and a diverse range of non-sequential features (e.g., the phenotype features) to make more accurate predictions of disease progression.
Using real-world patient medical data, we showed that the phenotype features can enhance the performance of a variety choice of model structures, including the revised Wide \& Deep model.
We envision that the clinical phenotype has potential in boosting model performance on other tasks, such as rare disease detection or pairing patients with matched clinicians or medications.

\end{document}